\documentclass[prx,twocolumn,aps,epsf,showpacs,superscriptaddress,longbibliography]{revtex4-1}
\usepackage[pdftex]{graphicx}

\usepackage{dcolumn}
\usepackage{bm}
\usepackage{epsfig}
\usepackage{latexsym} 
\usepackage{amsmath}
\usepackage{amssymb}
\usepackage{color}
\usepackage{array}
\usepackage{bbm}
\usepackage{hyperref}
\usepackage{color}
\usepackage{array}
\usepackage{cancel}
\usepackage{ulem}

\usepackage{booktabs}
\newcolumntype{C}[1]{>{\centering\arraybackslash}m{#1}}
\AtBeginDocument{
\heavyrulewidth=.08em
\lightrulewidth=.05em
\cmidrulewidth=.03em
\belowrulesep=.65ex
\belowbottomsep=0pt
\aboverulesep=.4ex
\abovetopsep=0pt
\cmidrulesep=\doublerulesep
\cmidrulekern=.5em
\defaultaddspace=.5em
}
\usepackage{booktabs}
\newcolumntype{R}[1]{>{\raggedleft\arraybackslash}p{#1}}
\AtBeginDocument{
\heavyrulewidth=.08em
\lightrulewidth=.05em
\cmidrulewidth=.03em
\belowrulesep=.65ex
\belowbottomsep=0pt
\aboverulesep=.4ex
\abovetopsep=0pt
\cmidrulesep=\doublerulesep
\cmidrulekern=.5em
\defaultaddspace=.5em
}

\newcommand{\<}{\langle}

\newcommand{\up}{\uparrow}
\newcommand{\down}{\downarrow}
\renewcommand{\t}{\tau}
\renewcommand{\>}{\rangle}
\renewcommand{\(}{\left(}
\renewcommand{\)}{\right)}
\renewcommand{\[}{\left[}
\renewcommand{\]}{\right]}
\renewcommand{\v}[1]{{\bf #1}} 

\renewcommand{\d}{\partial}

\newcommand{\eps}{\epsilon}
\newcommand{\p}{^{\prime}}
\renewcommand{\H}{\mathcal{H}}

\newcommand{\h}{\hat}
\renewcommand{\t}{\text}

\begin{document}
\title{Lattice Collective Modes
from a \\ Continuum Model of Magic-Angle Twisted Bilayer Graphene}
\author{Ajesh Kumar}
\email{These authors contributed equally to this work.}
\author{Ming Xie}
\email{These authors contributed equally to this work.}
\author{A. H. MacDonald}
\affiliation{Department of Physics, University of Texas at Austin, Austin, TX 78712, USA}
\begin{abstract}
We show that the insulating states of magic-angle twisted bilayer graphene support a 
series of collective modes corresponding to local particle-hole excitations on triangular lattice sites.
Our theory is based on a continuum model of the magic angle flat bands.
When the system is insulating at moir\'e band filling $\nu=-3$, our calculations show that 
the ground state supports seven low-energy modes that lie
well below the charge gap throughout the moir\'e Brillouin zone, one of which couples strongly to THz photons.
The low-energy collective modes are faithfully described by a model with a local $SU(8)$ 
degree of freedom in each moir\'e unit cell that we identify as the direct product 
of spin, valley, and an orbital pseudospin.  
Apart from spin and valley-wave modes, the collective mode spectrum includes a low-energy 
intra-flavor exciton mode associated with transitions between flat 
valence and conduction band orbitals.
\end{abstract}
\maketitle


\section{Introduction}
The insulating phases~\cite{Cao2018,Yankowitz1059,Sharpe2019,Lu2019,choi2019electronic,xie2019spectroscopic,jiang2019charge,kerelsky2019maximized} that occur at integer moir\'e band filling $\nu=nA_M$ 
in magic-angle twisted bilayer graphene (MATBG) are unusually rich. (Here $n$ is carrier density and $A_M$ is 
moir\'e pattern unit cell area.) Depending on the value 
of $\nu$ they can be translationally invariant
Chern insulators~\cite{Haldane1988,Chang2013,Sharpe2019,Serlin2020,Chen2020,nuckolls2020strongly,Zhang2019,Zhu2020,zhang2019nearly,bultinck2020mechanism,liu2019quantum,lian2020tbg,kang2020nonabelian,zhang2020correlated}
or unusual commensurate magnetization-density-wave states\cite{Bultinck2020,Liu2020correlated,ochoa2020strain}.
The insulating states are always in a gate-proximity-dependent competition with metallic states that 
can be superconducting\cite{Cao2018b,Lu2019,stepanov2020untying,saito2020independent,liu2020tuning}.
The proximity of superconducting and insulating states is reminiscent of the behavior  
of high-temperature superconductors, in which the nature of the insulating state is 
often thought\cite{HighTcLee,HighTcKeimer,HighTcProust} to be important in explaining superconductivity. 
The goal of the present work is to advance understanding of the properties of MATBG insulators 
by studying their low-energy particle-hole (p-h) collective excitations.

\begin{figure*}[t]
\centering
\includegraphics[width=\textwidth]{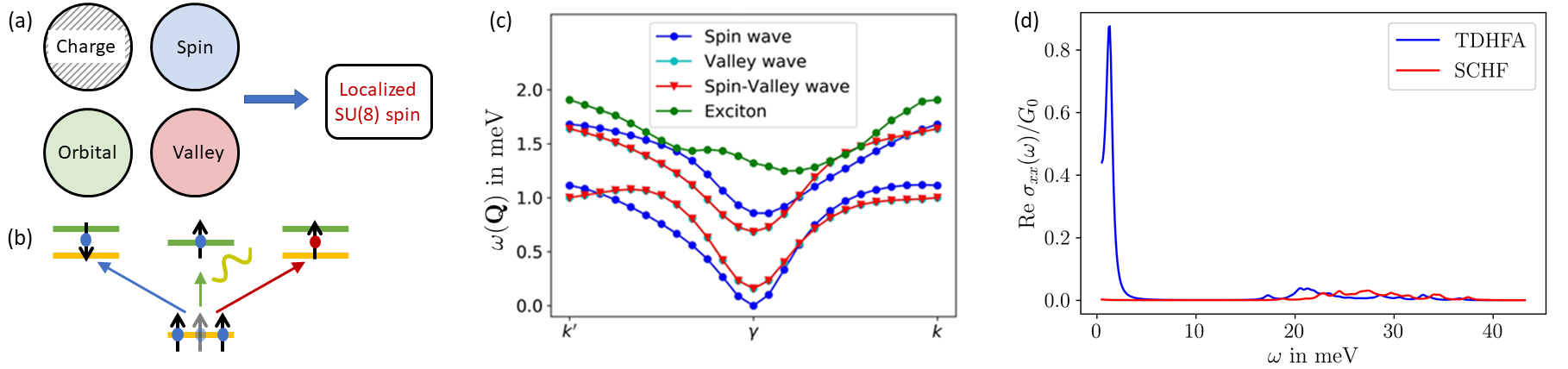}
\caption{(a) The charge degree of freedom is frozen at low energies allowing an effective local $SU(8)$ spin-$\times$-valley-$\times$-orbital
desciption. (b) Active band schematic illustrating various collective modes. The spin of the electrons is denoted by arrows, valley $+/-$ by blue/red filling, and valence/conduction bands by yellow/green respectively. The mean-field energies in the $-$ valley
are spin-degenerate, and therefore, only spin $\up$ is depicted. The intra-flavor exciton mode consists of transitions to the 
dressed conduction band of the occupied spin/valley, whereas the spin-wave and valley-wave modes include transitions to two empty bands. The intra-flavor mode couples strongly to light which is indicated by the wavy line.
(c) Spectra of all the low-energy collective modes versus mode momentum $\v{Q}$. (d) A comparison of the real part of the optical longitudinal conductivity calculated within TDHFA and directly using the SCHF bands (see Appendix~\ref{app:conductivity}). $G_0=2e^2/h$ is the conductance quantum. The TDHFA plot shows a significant peak due to the intra-flavor mode in the THz frequency range.}
\label{fig:coll_schematic}
\end{figure*}

The traditional approach to understanding the collective excitations of an insulator 
is to start from a lattice model in which electrons can occupy one of a small set of
atomic-like Wannier states associated with each lattice site.  In the simplest case of a 
one-band Hubbard model, for example, only two-spin states are available on each lattice site. 
That approach is not available in the MATBG
case because of the topological properties~\cite{Song2019,zou2018band,Po2019,Liu2019,ahn2019,song2020tbg}
of MATBG's valley-projected bands, 
which present an obstruction to the identification of useful Wannier orbitals.  
Instead one must either start from a model with many Wannier orbitals for each MATBG flat-band
or from a continuum model that does not restrict orbital wavefunctions.  
In this work, we develop a continuum model theory\cite{Bistritzer2011} of the low-energy collective 
particle-hole excitations that can be applied to any insulating state and is similar 
to the time-dependent mean-field approaches used previously\cite{Wu2020,Alavirad2019,Kwan2020} to 
calculate the spin-wave excitations of spin-polarized insulating states.
The continuum model approach has the incidental
advantage that it is possible to account for long-range Coulomb interactions
without truncation \cite{Xie2020}.  
We focus specifically on the insulating state at $\nu=-3$ carriers per moir\'e unit cell where 
the insulating ground state is fully spin and valley polarized and all band degeneracies are lifted 
by the broken spin and valley symmetry~\cite{Xie2020}. We note that our results apply for the $\nu=3$ insulator as well which is more commonly seen experimentally~\cite{Cao2018,Yankowitz1059,Sharpe2019,Lu2019}, since the model we study is nearly particle-hole symmetric~\cite{Bistritzer2011,Song2019}. We retain both both valence and conduction flat bands,
and show that this is essential to capture qualitative aspects of the excitation spectrum.

The main question we set out to answer is this.  Are there collective excitations that lie below the continuum of 
inter-band particle-hole excitations throughout the moir\'e pattern's Brillouin-zone?  
Any such collective excitation is expected to act like a bosonic degree of freedom that lives on the 
moir\'e pattern's triangular lattice.  These modes, if present, will control the dynamical response and thermal 
fluctuation properties of the insulating states at energies below their gaps.
The ground state spin and valley pseudo-spin quantum numbers of the $\nu=-3$ insulating state 
are $S_z=N/2$ and $T_z=N/2$, with $N$ being the total electron number.  By explicit calculation we find seven collective excitations 
with zone-boundary energies $\sim 2$ meV, well below the threshold for 
interband particle-hole excitations at $\sim 10$ meV: two spin-waves ($S_z=-1,T_z=0$), two spin-valley waves
($S_z=-1,T_z=-1$), two-valley waves ($S_z=0,T_z=-1$) and one intra-flavor exciton mode ($S_z=0,T_z=0)$.
To a very good approximation, these excitations correspond to particle-hole pairs 
localized within the same moir\'e unit cell, and correlated across the 2D system.  
Even though it is not possible to use a Wannier-function lattice 
model to describe the fermionic charged quasiparticles of the insulating state, the collective excitations map to 
those of a triangular lattice model with an eight-fold spin-$\times$-valley-$\times$-orbital degree of freedom 
on each site that is analogous to the $SU(8)$ degree-of-freedom within the $N=0$ Landau level of 
Bernal bilayer graphene.   The orbital degree-of-freedom in MATBG is connected to the property that flat bands always occur in 
conduction/valence pairs, but is more complicated than the orbital degree of freedom in Bernal quantum Hall bilayers,
because both conduction and valence band continuum model spinors exhibit complex wavevector-dependent 
entanglement between bilayer sublattice and orbital degrees of freedom.  


\section{Active-band projected self-consistent Hartree-Fock}

Before beginning our description of the details of the calculation, we comment 
on the symmetries of MATBG and on the residual symmetries of the self-consistent Hartree-Fock (SCHF) ground state. 
The symmetries satisfied by the continuum-model band Hamiltonian 
are six-fold rotational symmetry ($C_6$), $y \leftrightarrow -y$ mirror symmetry ($M_y$), time-reversal invariance,
and separate charge conservation and spin-rotational symmetry in the two valleys. 
Long-range Coulomb interactions satisfy $SU(4)$ spin-valley invariance. 
The total internal symmetry is therefore $U(2)_+ \times U(2)_-$, where $+$ and $-$ label the two valleys.
The mean-field ground state at $\nu=-3$
breaks time-reversal symmetry and $U(2)_+$ down to $U(1)_+ \times U(1)_+$ to achieve a fully spin and valley polarized insulating state without breaking the moir\'e pattern's translational symmetry, but does break $C_3$ rotational symmetry~\cite{Xie2020}.
In order to faithfully capture the low-energy collective modes of this ferromagnet,
considering for practical reasons only particle-hole excitations within the 
active flat-band subspace, we must project the self-consistent Hartree-Fock mean-field ground state 
calculation onto the active bands as well.  
This point is discussed explicitly for the Goldstone spin-wave mode in Appendix~\ref{app:gapless},
and is a reasonable approximation for the interaction strengths that we consider. 
We then calculate collective excitations, restricting to the eight (two per flavor) active SCHF bands 
retained employing the time-dependent Hartree-Fock approximation (TDHFA) (also known as the generalized random phase 
approximation GRPA)~\cite{Negele1982,Pines1994}.

The starting point of our study is a continuum model~\cite{Bistritzer2011} single-particle Hamiltonian 
$\H_{BM}$ that describes the twisted bilayer
in terms of spatially periodic sublattice-dependent local tunneling between graphene layers with Dirac-cone spectra.  
Following Refs.~\cite{Nam2017,Carr2019}, we incorporate the effects of lattice relaxation on the electronic band 
structure by taking the ratio of the strength of the inter-layer hopping between AA sites and AB sites to be $0.8$. 
The resulting electronic structure isolates two relatively flat low-energy active bands (labeled $A$ below)
from a set of well-separated higher energy remote ($R$) bands.
The active-band projected SCHF calculation is then conveniently described in the eigenbasis of the continuum model 
Hamiltonian $\H_{BM}$: $|\psi_{\mu,m,\v{k}} \>$, where $\mu$ is flavor (spin and valley), $m \in A/R$ is band,
and $\v{k}$ is momentum in the moir\'e Brillouin zone. 
We seek single-Slater-determinant (HF) states with one filled active band, which corresponds to $\nu=nA_M=-3$
since neutrality is reached when four of the eight active bands are occupied.

Assuming that the ground state does not mix flavors, the band density-matrix for 
flavor $\mu$ and wavevector $\v{k}$ is 
\begin{align}
\rho_{m,n}(\v{k},\mu) = \sum_i {z^i}^*_{\mu,m,\v{k}} z^i_{\mu,n,\v{k}}
\end{align} 
where $z^i_{\mu,m,\v{k}}$ is the quasiparticle wavefunction for band $i$, and 
$i$ is summed over all occupied bands with flavor $\mu$, 
including remote bands whose wavefunctions are frozen at their single-particle values.
It follows that the the Hartree-Fock self-energy 
\begin{align}
\Sigma_{\mu,m;\mu,n}(\v{k}) = \frac{1}{A_s} \sum_{\v{q},\sigma} \sum_{i,j \in {R/A}} \left\lbrace V^H_{minj}(\v k, \mu; \v q, \sigma) \rho_{i,j}(\v{q, \sigma}) - \right.  \nonumber\\
\left. - V^F_{minj}(\v k, \mu; \v q, \sigma) \rho_{i,j}(\v{q, \sigma}) \right\rbrace,
\end{align}
where $A_s$ is the finite-size system area,
and $V^H_{minj}(\v k, \mu; \v q, \sigma) = \<\psi_{\mu,m,\v{k}}, \psi_{\sigma,i,\v{q}} | \hat{V} | \psi_{\mu,n,\v{k}}, \psi_{\sigma,j,\v{q}} \>$ and $V^F_{minj}(\v k, \mu; \v q, \sigma) = \<\psi_{\mu,m,\v{k}}, \psi_{\sigma,i,\v{q}} | \hat{V} | \psi_{\sigma,j,\v{q}}, \psi_{\mu,n,\v{k}} \>$ are two-particle matrix elements calculated using the the long-range Coulomb interaction $\hat{V}$.
Since we assume that interaction dressing is included in the Dirac Hamiltonian of isolated neutral graphene sheets
(with density matrix $\rho_{\t{iso}}$), we regularize the self-energy by setting $\rho \rightarrow \rho - \rho_{\t{iso}}$. 
When frozen, the remote bands contribute a term $\Sigma^R$ to the HF self-energy that 
acts as a band-mixing external field.  
The HF self-energies are diagonal in flavor ($\mu$) because of the flavor independence ($SU(4)$ invariance) of 
the long-range Coulomb interactions and the flavor-diagonal nature of the SCHF ground state.
The two-particle matrix elements are evaluated from the plane-wave expansions of the band wavefunctions.
In the plane-wave basis, the two-dimensional Coulomb interaction is $V(\v q) = 2 \pi e^2/\epsilon q$ 
where $\v q$ is the momentum transfer, and $\epsilon$ is an effective dielectric constant. 
For the explicit illustrative calculations presented below we set the twist angle $\theta = 1.1^{\circ}$
and choose $\epsilon^{-1} = 0.06$.  (It is important to recognize that the appropriate values of these two parameter are
device-dependent.) 

\begin{figure}
\centering
\includegraphics[height=3in]{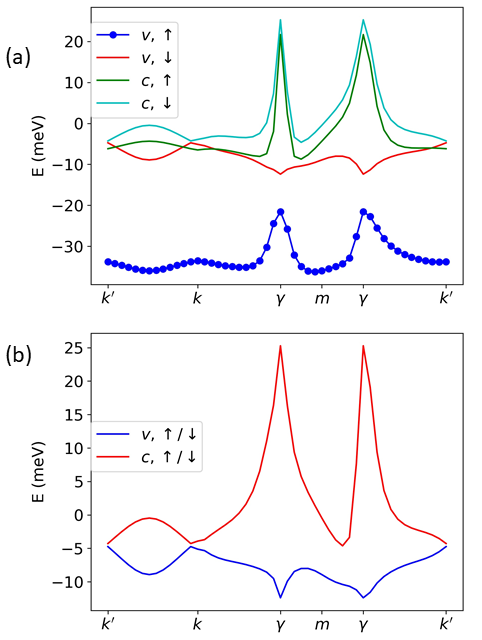}
\caption{Active-band projected Hartree-Fock band structure. (a) Valley +: Filled bands are indicated by filled circles. (b) Valley -: Bands are empty and spin-degenerate. Time-reversal symmetry breaking gaps the Dirac points in the filled flavor sector and 
yields bands with non-zero Chern numbers.} 
\label{fig:bands} 
\end{figure}

The SCHF eigenstates are then obtained by self-consistently diagonalizing the Hamiltonian:
\begin{align}
\H_{HF}[\rho] = \H_{BM} + \Sigma^R + \Sigma^A[\rho]
\end{align}
Fig. \ref{fig:bands} shows a plot of the SCHF band-structure at $\nu=-3$, 
for the fully spin and valley polarized ground state. 
For our subsequent analyses, we define the fermion annihilation operator for the active SCHF states as $f_{\mu,b,\v{k}}$, where $\mu$ denotes flavor, and $b=c,v$ is a band index with $v$ referring to the occupied lower energy state and $c$ 
referring to one of the higher energy unoccupied states.

\section{Collective modes}
With the SCHF band-structure in hand, we now turn to the computation of collective modes. First, we describe our method of computation in general, and then focus on the collective excitations within the active-band subspace illustrated in Fig. \ref{fig:coll_schematic}. 

We look for collective excitations with momentum $\v Q$ over the ground state $|0 \>$ of the form:
\begin{align}
\sum_{I,\v{k}}\( u^I_{\v{k}}(\v{Q}) f^{\dagger}_{p(I),\v{k+Q}} f_{h(I),\v{k}} + v^I_{\v{k}}(\v{Q}) f^{\dagger}_{h(I),\v{k}} f_{p(I),\v{k}-\v{Q}}\) |0 \>
\end{align}
where $I$ is an excitation label, and we have combined the flavor index with the band index by using $p(I)$ and $h(I)$ to label particle (empty) and hole (filled) states corresponding to the excitation label $I$. Here, $|0\>$ is the ground state of the system which includes one and two p-h fluctuations over the SCHF ground state. Even though its precise form is not known {\it a priori}, we can determine the amplitude of p-h excitations $u(\v{Q})$, and de-excitations $v(\v{Q})$, starting from $|0 \>$ within TDHFA. Large $\left|v(\v{Q})/u(\v{Q})\right|$ is indicative of strong p-h fluctuations in $|0\>$. In our calculations, we find it to be small ($\lessapprox 0.015$). The coefficients $u$, $v$, and the energy of the collective mode $\omega(\v{Q})$ are obtained by solving the non-Hermitian eigenvalue problem (Appendix~\ref{app:coll_modes} provides a derivation from linear response theory):
\begin{align}
&\sum_{J,\v{k\p}} A^{IJ}_{\v{k},\v{k\p}}(\v{Q}) u^J_{\v{k\p}}(\v{Q}) + B^{IJ}_{\v{k}, \v{k\p}}(\v{Q}) v^J_{\v{k\p}}(\v{Q}) \nonumber \\
&= \(\omega(\v{Q}) - \Delta^I_{\v{k}}(\v{Q}) \) u^I_{\v{k}}(\v{Q}) \nonumber \\
&\sum_{J,\v{k\p}} \({A^{IJ}_{\v{k},\v{k\p}}}(\v{-Q})\)^* v^J_{\v{k\p}}(\v{Q}) + \(B^{IJ}_{\v{k}, \v{k\p}}(\v{-Q})\)^* u^J_{\v{k\p}}(\v{Q}) \nonumber \\
&= -\(\omega(\v{Q}) + \Delta^I_{\v{k}}(\v{-Q}) \) v^I_{\v{k}}(\v{Q})
\label{eq:tdhf}
\end{align}
where $\Delta^I_{\v{k}}(\v{Q}) \equiv \eps_{p(I),\v{k+q}}-\eps_{h(I),\v{k}}$, $\eps_{b,\v{k}}$ 
is a Hartree-Fock ground state eigenvalue, and $A$ and $B$ are matrix elements of the Coulomb interaction in the HF eigenbasis: $A^{IJ}_{\v{k},\v{k\p}}(\v{Q}) \equiv \< \Omega| f_{h(I),\v{k}} f_{p(J),\v{k\p}+\v{Q}}  \hat{V}  f^{\dagger}_{p(I),\v{k}+\v{Q}} f^{\dagger}_{h(J),\v{k\p}}|\Omega \>$ and $B^{IJ}_{\v{k},\v{k\p}}(\v{Q}) \equiv \< \Omega| f_{h(I),\v{k}} f_{h(J),\v{k\p}}  \hat{V}  f^{\dagger}_{p(I),\v{k}+\v{Q}} f^{\dagger}_{p(J),\v{k\p}-\v{Q}}|\Omega \>$. Here $|\Omega \>$ is the vacuum state. The matrix $A$ couples a p-h (de-)excitation with a (de-)excitation, and $B$ couples a p-h excitation with a de-excitation. 
The above equation can be recast in a concise way as,
\begin{align}
    \begin{bmatrix}
    A(\v{Q})+\Delta(\v{Q})   &B(\v{Q})  \\
    -B(-\v{Q})^* &-A(-\v{Q})^*-\Delta(-\v{Q})
    \end{bmatrix}&
    \begin{bmatrix}
    u(\v{Q})   \\
    v(\v{Q})
    \end{bmatrix}\nonumber \\
    = \omega(\v{Q})&
    \begin{bmatrix}
    u(\v{Q})   \\
    v(\v{Q})
    \end{bmatrix}
    \label{eq:tdhf_compact}
\end{align}
where $\Delta$ is present only in the diagonal. The matrices $A(\v{Q})$ and $B(\v{Q})$ satisfy: $A^{IJ}_{\v{k},\v{k\p}}(\v{Q}) = \(A^{JI}_{\v{k\p},\v{k}}(\v{Q})\)^*$, $B^{IJ}_{\v{k},\v{k\p}}(\v{Q}) = B^{JI}_{\v{k\p},\v{k}}(-\v{Q})$; Eq.~(\ref{eq:tdhf_compact}) thus represents a non-Hermitian eigenvalue problem. However, the collective mode energies $\omega(\v{Q})$ are guaranteed to be real, as we now show. 

We discuss in Appendix~\ref{app:stability} that the matrix
\begin{align}
    \begin{bmatrix}
    A(\v{Q})+\Delta(\v{Q})   &B(\v{Q})  \\
    B(-\v{Q})^* &A(-\v{Q})^*+\Delta(-\v{Q})
    \end{bmatrix}
    \equiv S(\v{Q}) = S(\v{Q})^{\dagger}
\end{align}
is the Hessian of the total SCHF ground state energy functional $E[\rho] = \text{Tr}(\H_{HF}\rho)$, 
which is positive definite for a converged SCHF solution. Therefore it admits a Cholesky decomposition: 
$S = TT^{\dagger}$, where $T$ is a lower triangular matrix. 
We can then map Eq.~(\ref{eq:tdhf_compact}) to a Hermitian eigenvalue problem by defining the eigenvector $\xi^T(\v{Q}) \equiv \begin{bmatrix} u(\v{Q}), &v(\v{Q}) \end{bmatrix}$, and linear operators $X$ and $Z$ that are Pauli matrices acting on the collective excitation/de-excitation degree-of-freedom:
\begin{align}
T(\v{Q})^{\dagger} Z T(\v{Q}) \( T(\v{Q})^{\dagger} \xi(\v{Q})\) = \omega(\v{Q}) \( T(\v{Q})^{\dagger} \xi(\v{Q})\)
\end{align}
Thus, the problem reduces to diagonalizing the Hermitian matrix $T^{\dagger}(\v{Q}) Z T(\v{Q})$. In this notation, the non-Hermitian problem is: $Z S(\v{Q}) \(\xi(\v{Q})\) = \omega(\v{Q}) \(\xi(\v{Q})\)$. 
This solution, however, introduces an artificial doubling of the Hilbert space of collective excitations, analogous to the particle-hole redundancy in the Bogoliubov-de Gennes formalism: The eigenvalues of $T^{\dagger}(\v{Q}) Z T(\v{Q})$ occur in positive and negative pairs at momenta $\v{Q}$ and $\v{-Q}$, since $XZS(\v{Q}) = -ZS^*(\v{-Q})X$. For a given $\v{Q}$, we thus keep only the positive eigenvalues.

We now apply this method to calculate the excitation spectra of MATBG insulators.
Since total spin ($S_z$) and valley ($T_z$) components along the polarization direction are conserved quantities, 
we can Bloch diagonalize $T^{\dagger}(\v{Q}) Z T(\v{Q})$ and classify excitations 
depending on the $S_z$ and $ T_z$ quantum numbers of the collective modes.

\subsection{Intra-flavor modes}
\begin{figure}
\centering
\includegraphics[height=2.3in]{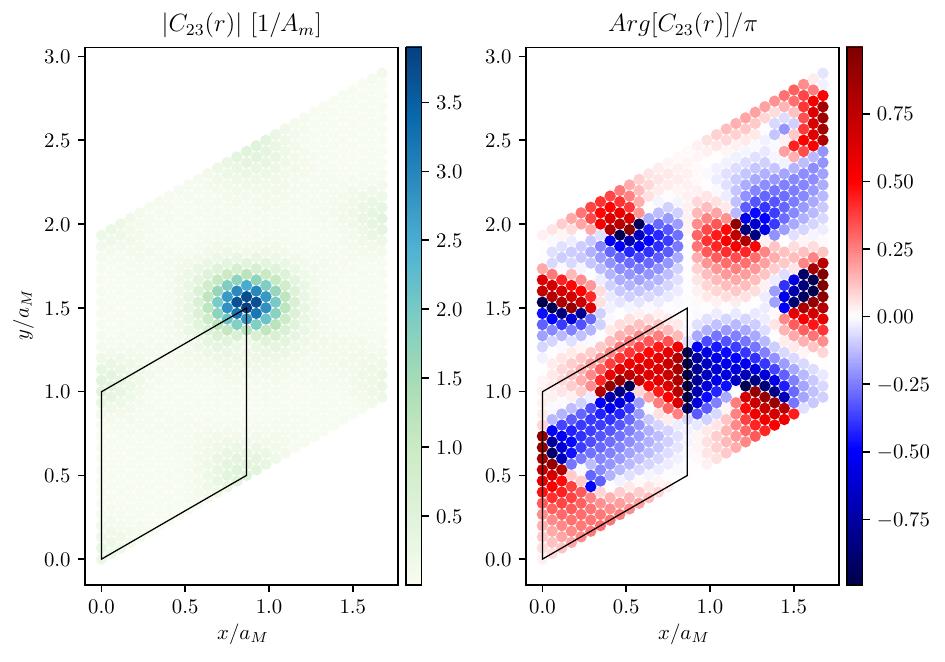}
\caption{The $C_{23}$ component of the real space wavefunction of the intra-flavor mode. 
The left and right panels plot the magnitude and the phase of $C_{23}$ as a function of electron-hole center-of-mass position.
The black solid lines mark a moir\'e unit cell with AA sites on its corners.
} 
\label{fig:realspacewf} 
\end{figure}

We first consider collective excitations of the form
\begin{align}
\sum_{\v{k}} &\left( u_{\v{k}}(\v{Q}) f^{\dagger}_{(+,\up,c),\v{k+Q}} f_{(+,\up,v),\v{k}} \right. \nonumber \\ 
&\left. + v_{\v{k}}(\v{Q}) f^{\dagger}_{(+,\up,v),\v{k}} f_{(+,\up,c),\v{k}-\v{Q}} \right) |0 \>
\end{align}
which preserve flavor.  For $\nu=-3$ the intra-flavor modes with wavevector $\v{Q}$ are constructed from 
transitions between the occupied active band states at wavevector $\v{k}$ and the unoccupied states of the same flavor at wavector $\v{k+Q}$.
The number of excitations with a given wavevector $\v{Q}$ is equal to the number of $\v{k}$'s in the moir\'e Brillouin-zone,
{\it i.e.} it is equal to the number of moir\'e unit cells in the system.  As we discuss below, only 
one of these modes has an energy that is well below the inter-band particle-hole continuum.
We can convert these wavefunctions, evaluated at zero electron-hole separation,
from a wavevector $\v{Q}$ representation to a real space representation 
by Fourier transforming with respect to excitation wavevector: 
\begin{align}
	\Psi^{ex}_{\alpha\beta}(\v{r},\v{R}) &= 
	\sum_{\v{k},\v{Q}} u_{\v{k}}(\v{Q}) e^{-i\v{Q} \cdot \v{R}} 
	   \psi^*_{c, \alpha, \v{k}+\v{Q}}(\v{r}) 
	   \psi_{v, \beta,\v{k}}(\v{r}),
\end{align}
where $\v{R}$ is a triangular lattice vector, 
$\psi_{c/v, \alpha, \bm{k}}(\bm{r})=\langle \alpha, \bm{r}|\psi_{(+,\up,c/v),\v{k}}\rangle$, and $\alpha$, $\beta$ are 
combined sublattice$\times$layer indices.
$\Psi^{ex}$ can be decomposed by Pauli matrices into sublattice and layer dependent contributions:
$C_{sl}=\sum_{\alpha\beta}(\sigma_{s}\tau_{l})_{\alpha\beta} \Psi^{ex}_{\alpha\beta}(\bm{r})$.
Fig.~\ref{fig:realspacewf} plots the coefficient, $C_{23}$, which has the largest weight, of the $\v{R}=0$ wavefunction.
The most important property of this excitation is that it is localized within one unit cell of the moir\'e pattern.
The low-energy mode is constructed from correlated local rotations in the $SU(2)$ orbital space on 
different lattice lattice sites.  These excitation energies are much smaller than the charge gaps of the insulator,
the energy needed to add distant uncorrelated electron-hole pairs, because the latter include the energy cost 
of doubly occupying a moiré unit cell, whereas the former do not.  

\subsection{Spin- and valley-waves}
\begin{figure}
\centering
\includegraphics[height=2.5in]{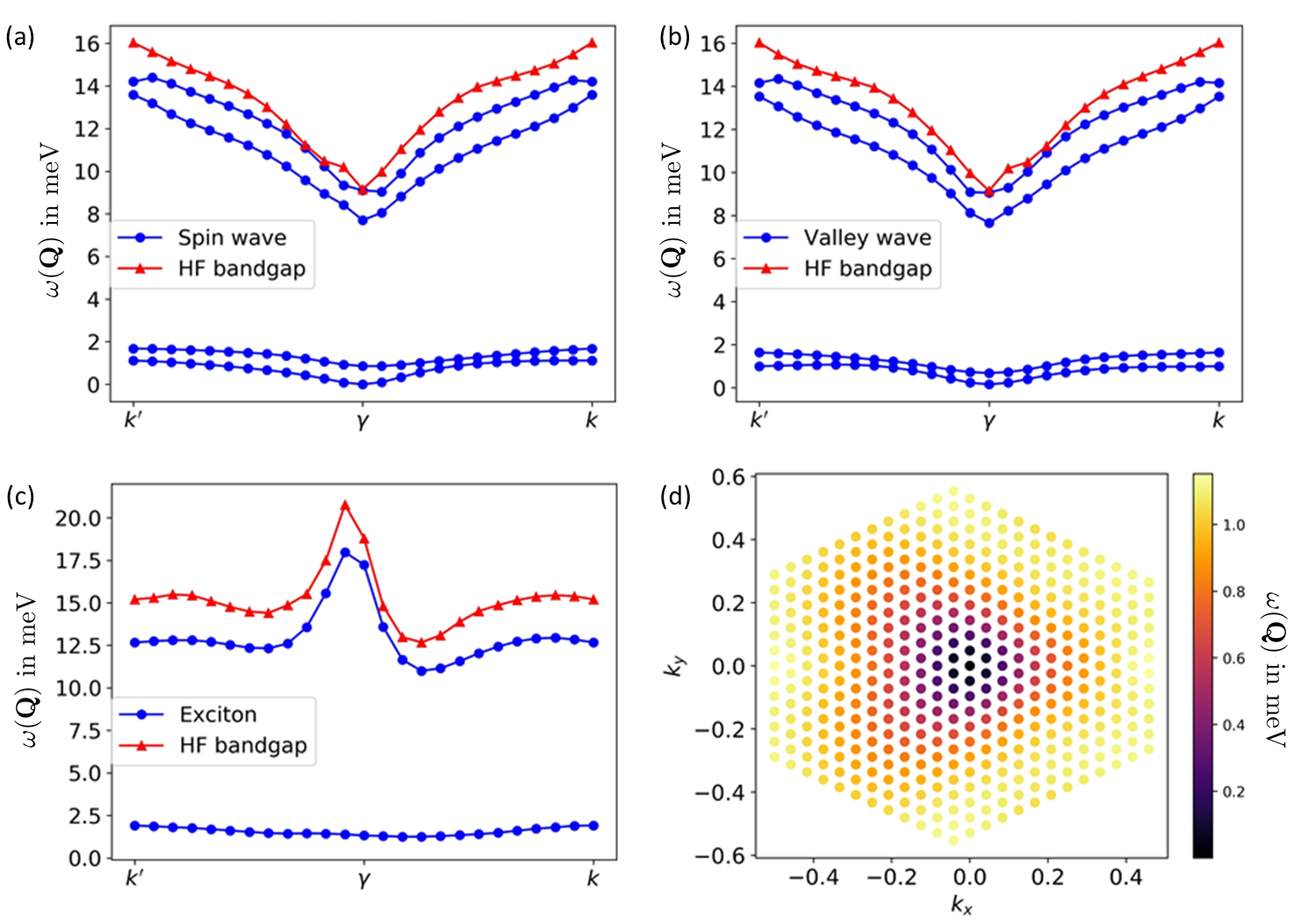}
\caption{(a-c) Comparison of low-energy collective excitation energies with the Hartree-Fock mean-field
bandgaps show a large separation of energy scales.  The unoccupied mean field bands are shifted by 
interactions with the occupied states, but these energies are qualitatively reduced when an electronic state 
is rotated within its $SU(8)$ space, instead of being added to the system.
(d) A $2d$ plot of the lowest-energy gapless spin-wave spectrum is mirror symmetric but breaks $C_3$ symmetry. $k_x$ and $k_y$ are shown in units of the moir\'e reciprocal lattice vectors. }
\label{fig:coll_gaps}
\end{figure}

We now solve for spin-waves within the $+$ valley, and valley-waves within the spin-$\up$ sector. 
Since there are no hole states with spin $\downarrow$ and valley $-$ in the SCHF ground state, these collective modes do not contain p-h de-excitations. This can also be seen by noting that the matrix $B=0$ for these cases, due to the $SU(4)$ symmetry of long-range Coulomb interactions. The spin-wave collective modes therefore, take the form:
\begin{align}
\sum_{\v{k}}&\( u^1_{\v{k}}(\v{Q}) f^{\dagger}_{(+,\down,v),\v{k+Q}} f_{(+,\up,v),\v{k}} \right. \nonumber \\
&\left.+ u^2_{\v{k}}(\v{Q}) f^{\dagger}_{(+,\down,c),\v{k+Q}} f_{(+,\up,v),\v{k}}\) |0 \>
\label{eq:spin_wave}
\end{align}
The calculation of the coefficients $u$ and the spin-wave energy $\omega(\v{Q})$ using Eq.~(\ref{eq:tdhf}) now turns into a Hermitian eigenvalue problem (with no redundant solutions):
\begin{align}
\sum_{J,\v{k\p}} A^{IJ}_{\v{k},\v{k\p}}(\v{Q}) u^J_{\v{k\p}}(\v{Q}) = \(\omega(\v{Q}) - \Delta^I_k(\v{Q}) \) u^I_{\v{k}}(\v{Q})
\label{eq:spin_wave_eqn}
\end{align}
The valley-wave and the spin-valley-wave calculations proceed analogously; we diagonalize the corresponding $A$ matrices to obtain the collective mode dispersions shown in Fig. \ref{fig:coll_gaps}.  Spin-wave, valley-wave, and spin-valley wave modes are all doubled because of the 
additional orbital degree of freedom present in the low-energy Hilbert space.  

\section{Discussion}

In MATBG eight low-energy flat bands are spectrally isolated from higher energy bands.
The many-electron Hilbert space is therefore closely analogous to that of the Bernal bilayer graphene
in a strong magnetic field where the $N=0$ Landau level has an eight-dimensional vector space
available to electrons at each guiding center~\cite{mccann2006landau}.  In both cases the eight-dimensional space is the direct 
product of spin, valley, and an additional two-dimensional space spanned by two orthogonal spinors with
components on the four sublattices of the honeycomb bilayer.  In the $N=0$ Landau level 
case the two spinors are very simple - they are localized almost entirely on a single sublattice and 
have the orbital structure of either $n=0$, or $n=1$, free-particle Landau levels~\cite{mccann2006landau}.  Experiments have shown
that insulating quantum Hall states occur at all integer Landau level filling factors $\nu_{LL}$, 
and also at many fractional filling factors, between $\nu=-4$ and $\nu=4$~\cite{feldman2009broken,martin2010local,Weitz812}.  
The integer filling factor insulating 
ground states are particularly simple, and are well approximated by single-Slater-determinant states in which spin, valley, and 
the additional orbital degree-of-freedom are polarized~\cite{cote2010orbital,jung2011lattice,Barlas_2012}.    
The orbital degree of freedom in the moir\'e superlattice case is much more subtle since both valence and 
conduction band spinors have strongly entangled sublattice and orbital dependences that change with 
wavevector in the moir\'e band.  Insulating states can appear at all integer filling factors 
for $\nu \in (-4,4)$, but are more likely to appear at some $\nu$'s that at others.  
(The appearance or absence of particular insulating states seems to have a dependence on 
twist-angle and gate-proximity that seems consistent with simple Stoner criteria\cite{Xie2020}).
It is nevertheless true that experiments~\cite{rozen2020entropic,saito2020isospin,zondiner2020cascade,wong2020cascade}
are however beginning to paint a clear picture that when insulators do occur, their ground states are well approximated 
by single-Slater-determinant states in which spin, valley, and orbital pseudospins are polarized in a way that minimizes
the ground state energy.  Even though the flat bands do have some dispersion~\cite{Xie2020,Bultinck2020}, 
interactions play as important a role as the single-particle Hamiltonian in determining the $\v{k}$-dependent orbital polarization of the insulating states.
This manuscript addresses the collective excitations of these insulating states.  As in the 
quantum Hall bilayer case~\cite{yang2006collective}, the number of low-energy collective modes when $M=\nu+4$ bands are occupied is 
$M(8-M)$, corresponding to transitions between all occupied bands and all empty bands.   
Although we preform explicit calculations only for $M=1$, 
our main goal is to reach conclusions that are independent of twist-angle and interaction strengths in a 
particular device, and of the particular band filling factor and the spin, valley, and orbital polarizations of its
ground state.  Our main finding is that there is a single collective mode for each inter-band transition that 
remains well below the gapped particle-hole continuum throughout the Brillouin-zone.  This property implies,
as we show by calculating the center-of-mass wavefunctions of the excitations, that the excitations represent local 
changes in the spin/valley/orbital state within a given moir\'e unit cell that avoid changing the number of 
electrons per period in any unit cell.  

In our explicit calculations for $\nu=-3$, we find seven relatively flat (bandwidth $\lessapprox 1$ meV) low-energy collective modes below $\approx 2$ meV, 
which for the interaction strength used in these calculations 
is an order of magnitude smaller than the charge gap $\approx 10$ meV. 
Therefore, at very low energies, there are seven independent magnon-like degrees of freedom at each $\v{Q}$, 
or equivalently, for each moir\'e unit cell in real space.
We plot all the non-degenerate 
low-energy collective modes in Fig.~\ref{fig:coll_gaps} along with the lowest energy
single-particle transition energy for each $\v{Q}$. 
The property that all collective modes have real positive energies indicates the stability of the fully spin and valley polarized 
SCHF ground state for the model parameters considered here; 
it is not possible for the system to lower its energy in any of the collective p-h excitation channels.   
Valley polarization in insulating states at odd moir\'e band filling factors 
is consistent with the reported quantum anomalous Hall effect~\cite{Sharpe2019,Serlin2020}. 

Our results strongly suggests that the low-energy physics of the $\nu = -3$ correlated insulator is captured by an effective $SU(8)$ spin model 
having one 8-component generalized spin per moir\'e unit cell that incorporates real spin, valley and orbital degrees of freedom. 
Including the conduction active bands is essential to faithfully capture all the low-energy collective modes. The spin-wave spectrum consists of a gapless mode which is the Golstone mode corresponding to $SU(2)_+$ to $U(1)_+$ symmetry breaking.
The valley-wave spectrum has a small gap $\approx 0.2$ meV even though the interactions are $SU(4)$ spin-valley invariant because
the single-particle Hamiltonians of the two valleys are different, breaking the $SU(4)$
symmetry down to $SU(2)_+ \times SU(2)_- \times U(1)_v$. 
The $U(1)_v$ symmetry corresponds to the conservation of valley polarizaton - 
the difference between the numbers of electrons in the two valleys or total $T_z$, 
but the total valley-angular momentum is not conserved. This property
is responsible for the gap in the valley-wave spectrum.  The dispersion of the 
valley wave mode provides a measure of the sensitivity of energy to spatial configurations of the valley pseudospin, and 
hence an estimate of the temperature to which valley order can survive.  
The spin-valley-wave modes are degenerate with the valley-wave modes because the 
Hamiltonian is invariant under independent spin-rotations in either valley; electrons in one valley 
are insensitive to the spins of electrons in the other valley.  The generalized anisotropy energies of the $SU(8)$
degrees of freedom of insulating MATBG, are more complex when the orbital degrees of freedom is involved, although 
our calculations show that the energy scales of spin, valley, and orbital dependent interactions are similar.
Because the splitting between conduction and valence bands is small at most wavevectors in the moir\'e 
Brillouin-zone, its contribution to the localized state Hamiltonian plays the role of a weak external field that 
contributes to the $SU(8)$ anisotropy landscape.   

It is clear from experiment~\cite{zondiner2020cascade,wong2020cascade,rozen2020entropic,saito2020isospin, stepanov2020untying, wu2020chern, Ming2020Hall} 
that the broken spin and valley flavor symmetries that characterize MATBG insulating ground states persist through broad fractional
filling factor intervals.  The charge gaps that are common at fractional filling factors in the quantum Hall case 
are so far absent in MATBG experiments.  Instead experiments show robust two-dimensional metallic states in some regions 
of filling factor, and properties that remain obscure in some other regions.  We anticipate that the collective modes 
discussed in this paper, including the intraflavor inter-band excitionic collective mode, will remain sharp in the metallic 
state.  It remains to see if they are responsible for the strange metal behavior evident in the temperature 
dependence of the resistivity~\cite{polshyn2019large,cao2020strange} and for the superconductivity, which 
in most cases emerges from metals with broken spin/valley flavor symmetries.


{\it Note Added}: After this work was completed we learned of closely related work that provides a complementary 
point of view on the collective modes of MATBG~\cite{khalaf2020soft,vafek2020hidden,bernevig2020tbg}.

\section{Acknowledgements}
We acknowledge helpful conversations with Naichao Hu, Chunli Huang, Elaine Li, Lukas Linhart, Pawel Potasz, Ashvin Vishwanath and Nemin Wei. AK was supported by the National Science Foundation through the Center for Dynamics and Control of 
Materials: an NSF MRSEC under Cooperative Agreement No. DMR-1720595.  MX and AHM were supported by 
DOE BES grant DE- FG02-02ER45958.  

\bibliography{CollectiveModesbib}

\appendix
\onecolumngrid
\section{Collective modes within TDHFA from linear response}
\label{app:coll_modes}
Here we review linear response theory within the TDHFA, to a generic one-body time-periodic perturbation $\H_1(t)$, using which we will obtain Eq.~(\ref{eq:tdhf}) in the main text for the collective modes. The TDHFA is essentially that, under the external perturbation, the many-body wavefunction stays a Slater determinant at all times. The single-particle density matrix $\rho$ corresponding to a Slater determinant wavefunction satisfies,
\begin{align}
\rho^2 = \rho; \quad \t{Tr}\rho = N
\label{eq:slater}
\end{align}
where $N$ is the total number of particles. The time evolution within the TDHFA is given by~\cite{Negele1982}:
\begin{align}
i \d_t \rho &= \[\H_{HF}[\rho] + \H_1(t), \rho \]
\label{eq_tdhf}
\end{align}

Starting with the static SCHF ground state $\rho_0$, which satisifes $\[\H_{HF}[\rho_0], \rho_0 \] = 0$, we wish to study the linear response of the system to a small $\H_1(t)$, which is periodic with frequency $\omega$: $\H_1(t) = \H_1 e^{-i\omega t} + h.c$. Assuming that the response of the density matrix is linear in $\H_1(t)$, we can write:
\begin{align}
\rho(t) = \rho_0 + \rho_1 e^{i\omega t} + {\rho}^{\dagger}_1 e^{-i\omega t}
\end{align}

However, the constraints on $\rho(t)$ specified by Eq.~(\ref{eq:slater}), imply that $\rho_1$ introduces only p-h fluctuations over $\rho_0$ and not p-p fluctuations at leading order in the external perturbation: 
\begin{align}
\< \psi_{p(I)/h(I),\v{k}} | \rho_1 | \psi_{p(I)/h(I),\v{k\p}} \> &= 0 
\end{align}
In order to describe the dependence of $\rho_1$ on $\H_1$, we first define the matrix elements of $\rho_1$ and $\H_1$:
\begin{align}
u^I_{\v{k}}(\v{Q}) &\equiv \< \psi_{p(I),\v{k+Q}} | {\rho}^{\dagger}_1 | \psi_{h(I),\v{k}} \> \nonumber\\
v^I_{\v{k}}(\v{Q}) &\equiv \< \psi_{h(I),\v{k}} | {\rho}^{\dagger}_1 | \psi_{p(I),\v{k-Q}} \> \nonumber \\
f^I_{\v{k}}(\v{Q}) &\equiv \< \psi_{p(I),\v{k+Q}} | \H_{1} | \psi_{h(I),\v{k}} \> \nonumber \\
g^I_{\v{k}}(\v{Q}) &\equiv \< \psi_{h(I),\v{k}} | \H_{1} | \psi_{p(I),\v{k-Q}} \>
\label{eq_pert_HF_basis}
\end{align}
Using the TDHFA equation of motion (Eq.~(\ref{eq_tdhf})), to leading order in the external perturbation, we get the linear response equation:
\begin{align}
\(A^{IJ}_{\v{k},\v{k\p}}(\v{Q}) - \omega \delta_{\v{k},\v{k\p}} \delta_{I,J} \) u^J_{\v{k\p}}(\v{Q}) + B^{IJ}_{\v{k}, \v{k\p}}(\v{Q}) v^{J}_{\v{k\p}}(\v{Q}) &= - f^I_{\v{k}}(\v{Q}) \nonumber \\
\( {A^{IJ}_{\v{k},\v{k\p}}(-\v{Q})}^* + \omega \delta_{\v{k},\v{k\p}} \delta_{I,J} \) v^{J}_{\v{k\p}}(\v{Q}) + {B^{IJ}_{\v{k},\v{k\p}}(-\v{Q})}^* u^J_{\v{k\p}}(\v{Q}) &= -{g^I_{\v{k}}(\v{Q})}
\label{eq_lin_res}
\end{align}


We rewrite the linear response equation (Eq.~(\ref{eq_lin_res})) in a compact form:
\begin{align}
\left(
    \begin{bmatrix}
    A(\v{Q}) + \Delta(\v{Q})   &B(\v{Q})  \\
    B(-\v{Q})^* &A(-\v{Q})^* + \Delta(-\v{Q})
    \end{bmatrix}
    -\omega
    \begin{bmatrix}
    1   &0  \\
    0   &-1 \\
    \end{bmatrix}
    \right)
    \begin{bmatrix}
    u(\v{Q})   \\
    v(\v{Q})
    \end{bmatrix}
    = -
    \begin{bmatrix}
    f(\v{Q})   \\
    g(\v{Q})
    \end{bmatrix}
    \label{eq:linear_response_compact}
\end{align}
Now we define the response function $R(\omega, \v{Q})$ that relates the change in the density matrix to the external perturation, by inverting Eq.~(\ref{eq:linear_response_compact}):
\begin{align}
    \begin{bmatrix}
    u(\v{Q})   \\
    v(\v{Q})
    \end{bmatrix}
    = R (\omega, \v{Q})
    \begin{bmatrix}
    f(\v{Q})   \\
    g(\v{Q})
    \end{bmatrix}
    \label{eq_rho_response}
\end{align}
The poles of $R$ correspond to the collective excitations of the system, and they are precisely given by Eq.~(\ref{eq:tdhf_compact}) of the main text.

\subsection{Optical conductivity}
\label{app:conductivity}
Here we calculate the electric current in response to a weak, spatially uniform and time-periodic external electric field, within TDHFA. The electric current operator is given by $\h{J^i} = -e\frac{\d \H_{BM}}{\d k_i}$, and since the momentum dependence of $\H_{BM}$ purely comes from the Dirac part, $\h{J^i}$ is diagonal in flavor and momentum. Consider an electric field along the $x$ direction corresponding to a gauge potential $\v{A} = i \omega E_x \h{x}$. This couples to the current operator in the Hamiltonian giving the term:
\begin{align}
\H^1(t) = \h{J^x} A e^{-i\omega t}
\end{align}
Since $\h{J^x}$ is diagonal in momentum and flavor, it only couples SCHF states at the same momentum, and same flavor. Therefore, only the intra-flavor collective mode at $\v{Q}=0$ contributes to the system's response to the electric field. We drop the $\v{Q}$ label in the following description for simplicity. Defining the matrix elements of $\h{J^x}$ in the HF basis:
\begin{align}
{J}_{\v{k}} \equiv \< \psi_{(+,\up,c),\v{k}} | \h{J^x} | \psi_{(+,\up,v),\v{k}} \> 
\end{align} 

We now express the conductivity $\sigma_{xx} = \<\h{J}^x\>/E_x$ in terms of the current matrix elements. Introducing a decay rate $\gamma$ (that cause broadening of the peaks at the collective mode energies that are otherwise delta-functions), the real part of the conductivity is:
\begin{align}
\t{Re} \sigma_{xx}(\omega) = \frac{\gamma}{\omega} \sum_i \frac{\t{sgn}(\omega^i)}{ (\omega - \omega^i)^2 + \gamma^2} \sum_{\v{k},\v{k\p}}
\begin{bmatrix}
    {J}^{*}_{\v{k}}   &{J}_{\v{k}}
    \end{bmatrix}
    \begin{bmatrix}
    u^i_{\v{k}} (u^i_{\v{k\p}})^*  &u^i_{\v{k}} (v^i_{\v{k\p}})^*   \\
    v^i_{\v{k}} (u^i_{\v{k\p}})^* &v^i_{\v{k}} (v^i_{\v{k\p}})^* 	\end{bmatrix}
    \begin{bmatrix}
    {J}_{\v{k\p}}   \\
    {J}^{*}_{\v{k\p}}
    \end{bmatrix}
    \label{app:eq_real_cond}
\end{align}
where $\omega^i$ and $u^i$, $v^i$ are the eigenvalue and the eigenvector corresponding to the collective mode labeled by $i$ at $\v{Q}=0$. Fig.~\ref{fig:coll_schematic}(d) shows the optical conductivity calculated using Eq.~\ref{app:eq_real_cond} for $\gamma=0.4$ meV. 

For comparison, we also calculate $\t{Re}(\sigma_{xx})$ by assuming no interactions between the SCHF states. In this case the poles of the response function $R$ are at energies corresponding to energy differences between the SCHF states, and the conductivity simplifies to:
\begin{align}
\t{Re} \sigma_{xx}(\omega) = \frac{\gamma}{\Omega} \sum_{\v{k}} \frac{|{J}_{\v{k}}|^2}{ \(\omega - (\eps_{(+,\up,c),\v{k}} - \eps_{(+,\up,v),\v{k}})\)^2 + \gamma^2}
\label{app:eq_schf_conductivity}
\end{align}
A comparison of the conductivities calculated using Eqs.~\ref{app:eq_real_cond} and \ref{app:eq_schf_conductivity} is shown in Fig.~\ref{fig:coll_schematic}(d). The TDHFA calculation shows a strong peak at the lowest intra-flavor collective mode energy which is absent in the SCHF version since it does not incorporate the contribution from the collective modes. Detection of this intra-flavor collective mode which is in the THz frequency range, in spectroscopy experiments is a tantalizing prospect.

\section{Stability of SCHF ground states}
\label{app:stability}
Given a converged SCHF ground state $\rho_0$, here we derive the Hessian of the HF energy functional, and hence, desribe the quadratic fluctuations about the ground state. We do this for a generic system by labelling the SCHF eigenstates by a single index: $|i \>$, and thereby, suppressing momentum and band labels used in the main text. All matrix elements below are in the SCHF eigenbasis.

For an infinitesimal perturbation $\delta \rho$ about $\rho_0$, Eq.~(\ref{eq:slater}) implies,
\begin{align}
\(\rho_0 + \delta \rho \)^2 = \rho_0 + \delta \rho
\end{align}
and therefore,
\begin{align}
\rho_0 \( \(\delta \rho\)^2 + \delta \rho \)\rho_0 &= 0 \nonumber\\
\sigma_0 \( \(\delta \rho\)^2 - \delta \rho \)\sigma_0 &= 0
\label{eq:constr}
\end{align}
where $\sigma_0 = 1-\rho_0$ is a projector on the empty band subspace, since $\rho_0$ is a projector on the filled band subspace. Labelling a empty states by $p$ and filled states by $h$, Eq.~(\ref{eq:constr}) implies 
\begin{align}
\delta \rho_{h, h\p} &= -\sum_p \delta \rho_{h, p} \delta \rho_{p, h\p} \nonumber \\
\delta \rho_{p, p\p} &= \sum_h \delta \rho_{p,h} \delta \rho_{h,p\p}
\label{eq:constraint_pp}
\end{align}

The elements $\delta \rho_{p,h}$ are hence the leading order variations of the density matrix. We now expand the total HF energy functional $E[\rho]$ to quadratic order in $\delta \rho_{p,h}$. For a general $\rho$,
\begin{align}
E[\rho] &= \sum_{i} \epsilon_i \rho_{i,i} + \frac{1}{2} \sum_{ijkl} \( V_{ijkl} - V_{ijlk} \) \rho_{k,i} \rho_{l,j}
\label{eq:tot_energy}
\end{align}
where $\epsilon$ are the SCHF eigenstate energies and $V_{ijkl} \equiv \<i, j | \hat{V} | k, l \>$. Expanding $E[\rho]$ about $\rho_0$:
\begin{align}
&E[\rho_0 + \delta \rho] - E[\rho_0] \approx \sum_{ij} \(\frac{\delta E}{\delta \rho_{i,j}}\)_{\rho_0} \delta \rho_{j,i} + \sum_{ijkl} \(\frac{\delta^2 E}{\delta \rho_{i,j} \delta \rho_{k,l}}\)_{\rho_0} \delta \rho_{j,i} \delta \rho_{l,k}
\label{eq:energy_exp}
\end{align}
From Eq.~(\ref{eq:tot_energy}), we have: $\(\frac{\delta E}{\delta \rho_{p,h}}\)_{\rho_0} = \(\frac{\delta E}{\delta \rho_{h,p}}\)_{\rho_0} = 0$. Therefore, the first term in Eq.~(\ref{eq:energy_exp}) only gets contributions from $\delta \rho_{p,p}$ and $\delta \rho_{h,h}$ and the second term from $\delta \rho_{p,h}$ and $\delta \rho_{h,p}$. Using Eq.~(\ref{eq:constraint_pp}),
\begin{align}
\sum_i \epsilon_{i} \delta \rho_{ii} &= \sum_{ph} \(\epsilon_p - \epsilon_h\)\delta \rho_{p,h} \delta \rho_{h,p}
\end{align}
Adding this to the contribution from the interaction matrix elements in the second term of Eq.~(\ref{eq:energy_exp}) and using $\delta \rho_{p,h} = \delta \rho_{h,p}^*$, we get:
\begin{align}
E[\rho_0 + \delta \rho] - E[\rho_0] \approx \sum_{hp h\p p\p}\begin{bmatrix}
\delta \rho^*_{h,p} &\delta \rho^*_{p,h}
\end{bmatrix}	\begin{bmatrix}
\(A+\Delta\)_{ph p\p h\p} &B_{ph p\p h\p} \\
B^*_{ph p\p h\p} &\(A+\Delta\)^*_{ph p\p h\p}
\end{bmatrix}	\begin{bmatrix}
\delta \rho_{h\p, p\p} \\
\delta \rho_{p\p, h\p}
\end{bmatrix}
\end{align}
where $A_{ph p\p h\p} \equiv V_{ph\p p h\p} - V_{ph\p p\p h}$, $B_{ph p\p h\p} \equiv V_{pp\p h h\p} - V_{pp\p h\p h}$ and $\Delta_{ph p\p h\p} \equiv \(\epsilon_p - \epsilon_h\) \delta_{p,p\p} \delta_{h,h\p}$. Having obtained the Hamiltonian for quadratic fluctuations about the SCHF ground state, the condition for the stability of the SCHF ground state is that the matrix:
\begin{align}
S = \begin{bmatrix}
A+\Delta &B \\
B^* &A^*+\Delta
\end{bmatrix}
\end{align}
is positive-definite, which we make use of in the main text. 

\section{Gapless spin-wave}
\label{app:gapless}
The spin-wave spectrum contains a quadratically dispersing gapless mode, as expected for an $SU(2)$ ferromagnet. Because we choose the basis for the collective modes to be the active-band subspace, the existence of the gapless mode relies crucially on freezing the remote bands in our SCHF calculation, as we now show. 

We begin by deriving a necessary condition on the $\v{Q}=0$ gapless spin-wave (magnon) wavefunction within TDHFA. Physically, the zero-energy magnon mode corresponds to a uniform spin-rotation of all spins in the ground state, because of the $SU(2)_+$ symmetry. Separating the spin part $|\chi_s\>$ of the SCHF eigenstates: $|(+,s,b),\v{k} \> = |\psi_{(+,s,b),\v{k}}\> \otimes |\chi_s\>$, we perform an infinitesimal spin-rotation on the occupied spin-$\up$ states: $|\chi_{\up}\> \rightarrow |\chi_{\up}\> + \alpha |\chi_{\down}\>$, which corresponds to a change in the density matrix $\delta \hat{\rho}$ given by,
\begin{align}
\< h(I),\v{k} | \delta \hat{\rho} | p(I),\v{k} \> = \alpha \< \psi_{h(I),\v{k}} | \psi_{p(I),\v{k}} \> \equiv \alpha c^I_{\v{k}}
\end{align}
where $I$ is an excitation label for spin-flip excitations within the $+$ valley, as in Eq.~(\ref{eq:spin_wave}) of the main text. The HF self-energy $\Sigma$ corresponding to the modified density matrix can be written in terms of the matrix $A$ as
\begin{align}
\<h(I),\v{k} | \Sigma | p(I),\v{k} \> = \alpha \sum_{J,\v{k\p}} A^{IJ}_{\v{k},\v{k\p}}(\v{Q}=0) c^J_{\v{k\p}}
\end{align}
The HF Hamiltonian in the rotated spin basis to $O(\alpha^2)$:
\begin{align}
\<h\p(I),\v{k} | \H_{HF} | p\p(I),\v{k} \> = \alpha \( \Delta^I_{\v{k}}(\v{Q}=0) c^I_{\v{k}} + \sum_{J,\v{k\p}} A^{IJ}_{\v{k},\v{k\p}}(\v{Q}=0) c^J_{\v{k\p}} \)
\end{align}
The diagonal elements are unchanged and are the SCHF eigenvalues. Now we impose the condition that the off-diagonal elements of the HF Hamiltonian do not change as well because of the $SU(2)_+$ symmetry, which implies that $u^I_{\v{k}}(\v{Q=0}) = c^I_{\v{k}}$ is a spin-wave solution of Eq.~(\ref{eq:spin_wave_eqn}) of the main text with $\omega(\v{Q}=0)=0$. This result establishes that a uniform spin-rotation does indeed correspond to a zero-energy spin-wave mode, and that its wavefunction $c$ is given by the overlap between the spatial parts of the majority and the minority spin states. It relies on $SU(2)_+$ symmetry, and crucially, having the same basis for the possible spin-flip excitations in the calculation for the spin-waves within TDHFA and those associated with a uniform spin-rotation of the SCHF ground state. In our case, since we consider collective excitations only within the active-band subspace, it is necessary to freeze the remote bands in the SCHF calculation.

\end{document}